\documentclass[11pt]{article}

\usepackage[T1]{fontenc}        
\usepackage{newtxtext,newtxmath} 

\usepackage[numbers]{natbib}
\usepackage{authblk}
\usepackage[a4paper,
            left=1in,
            right=1in,
            top=1in,
            bottom=1in]{geometry} 

\usepackage{graphicx}
\usepackage{lipsum} 
\usepackage{authblk} 
\usepackage{multirow}
\usepackage{tablefootnote}
\usepackage{graphicx}
\usepackage{pifont}
\usepackage{booktabs}
\usepackage[colorlinks=true, linkcolor=black, urlcolor=black, citecolor=black]{hyperref}
\title{\textbf{Gendered Digital Financing Adoption and Women's Financial Inclusion in Pakistan}}

\author{Abdul Wadood Asim}
\author{Khansa Zafar}
\author{Muhammad Raees\thanks{Contact: raees.se@must.edu.pk}}

\affil{Mirpur University of Science and Technology, Mirpur, AJK, Pakistan}

\begin{document}

\maketitle

\begin{abstract}
Financial inclusion is a longstanding concern across underdeveloped communities, particularly for women.
However, there are limited data-driven measures to first quantitatively identify such concerns and second to inform policies. 
In this work, we examine the digital money service adoption and women’s financial inclusion in the context of Pakistan.
We use the nationally representative Global Findex data from the World Bank to analyze how mobile money usage, when moderated by phone ownership, internet access, and education, affects women’s access to formal financial services. 
Our findings show that women who adopt mobile money services have significantly higher odds of accessing formal financial systems. 
Findings also reveal nuanced insights: internet access does not significantly impact inclusion, highlighting the influence of socio-cultural constraints.
Despite the limitations of using cross-sectional data and the absence of qualitative dimensions, our study contributes empirical evidence on gendered digital finance adoption. The findings have important implications for policy, including the need for women-centric fintech design and digital literacy reforms to bridge the gender gap in financial inclusion.
\end{abstract}


\section{Introduction}
Access to financial systems is not normally distributed across genders and social statuses~\cite{razzaq2024determinants}. 
For instance, in developing countries, like Pakistan, the gap in financial inclusion between men and women remains significantly wide~\cite{agarwal2020gender}. 
According to the Global Findex Report~\cite{worldbank2021}, only 13 percent of women in Pakistan have access to banking, compared to 34 percent of men. 
Major factors, beyond cultural and social, contributing to this gap include restricted mobility, limited access to digital tools, and inadequate financial literacy~\cite{rozalinda2024economic}\cite{singh2017bridging}. There are various underlying factors that affect financial inclusion. For example, women's education level, control over household income, and social status are primary factors creating this disparity. Many women involved in home-based businesses or informal work lack direct access to earnings~\cite{ghosh2019gender}. 
Women also face barriers to digital finance and may not fully utilize the available services due to a lack of autonomy or knowledge~\cite{kishorstudy}. 
Therefore, the gap widens not only in access but also in the actual use of financial systems~\cite{razzaq2024determinants}.

There are challenges relating to digital services that also impact financial inclusion.
For instance, mobile phone ownership remains low among women in various communities in Pakistan~\cite{benyacoub2021etude}. 
Even with some access to mobile services, women mainly rely on informal financial systems like cash exchanges or savings groups~\cite{razzaq2024determinants}.
For example, even financially literate women report avoiding mobile banking due to fear of fraud and limited digital literacy~\cite{bucher2021fearless}. 
While education, employment, and income are expected to improve access to financial services, the effect has not been translated or captured effectively~\cite{mndolwa2020gender}.
This potentially requires structural reforms to enhance digital and inclusive access to financial services~\cite{goel2023trends} \cite{ejaz2023financial}. 
On the policy level, national initiatives such as the Financial Inclusion Strategy~\cite{sbp2015national} and the Banking on Equality (BoE)~\cite{sbp2020banking} are aimed at increasing women’s access to financial services~\cite{razzaq2024determinants} \cite{noreen2022role}.
These policies promote gender-disaggregated data collection to create a more inclusive environment~\cite{jamal2023women}. 
However, the effect of these policies, particularly the digital financial services and the required literacy around these programs, is often unmeasured.

The mobile services offer digital alternatives that can extend financial access to women and underserved populations~\cite{awuah2025digital} \cite{uddin2025influence}. 
Although these services improve access, limited research has explored the actual usage of mobile accounts by women~\cite{connectedwomen2015} \cite{islam2016women}. 
Mobile financial services could be transformative; however, the critical question remains: Are people using these tools, and do these tools lead to financial empowerment or merely increase account ownership? 
Hence, with this work, we aim to explore the research question: \textbf{Does access to mobile money service improve women’s financial inclusion in Pakistan?}

In this study, we explore the relationship between mobile money adoption and women’s financial inclusion in Pakistan using nationally representative data. 
We use data from the Global Findex Report~\cite{gfindex2021} to understand the relationship between access, literacy, and usage patterns.
We explore enabling factors that influence how women engage with mobile financial services. 
We hypothesize that access to digital tools increases financial access. 
Our analysis reveals that many women face barriers such as mobile ownership, internet access, and literacy, among others. 
We also recognize the limitations of self-reported access and the absence of usage metrics. 
Thus, our analysis contributes by identifying which literacy indicators matter most and how they correlate with financial inclusion outcomes.
Based on these findings, we provide recommendations for exploring more research into financial inclusion measures to provide policy guidance.

The rest of the paper is organized as follows: Section II provides a review of the related literature. We define our methodology and data analysis approach in Section III. Section IV provides the results and their discussion along with some recommendations for further research. We conclude our paper in Section V.

\section{Literature Review}
Financial inclusion ensures equitable access to formal financial systems for every individual, which is an essential indicator for economic growth. Beyond access to financial systems, it leads to empowerment, self-reliance, and independent decision-making. 
However, financial inclusion in Pakistan remains a persistent challenge, particularly for women, who face structural and socioeconomic barriers in accessing formal financial services such as bank accounts or loans. 
The gender gap is particularly evident in underdeveloped communities~\cite{razzaq2024determinants} \cite{khoso2025enhancing}, though men too struggle with low financial literacy and lack of access. 
Even when services are available, the complexity of banking procedures and the fear of fraud discourage women from using them~\cite{reynolds2023exploring}. 
Despite national policies on financial inclusion~\cite{sbp2015national} \cite{sbp2020banking}, many interventions lack evaluation, and their real impact remains uncertain~\cite{javed2024gender}. 
These issues signal the need for a better understanding of how financial services are used and, more importantly, what prevents their usage, from a data-driven perspective.

A key driver of financial exclusion is the lack of financial literacy, particularly among women. 
According to~\cite{zahid2023women}, financial literacy strongly influences access and usage of financial services. 
However, only 13 percent of women in Pakistan have bank accounts, and informal saving methods such as rotating savings committees are more common than formal ones~\cite{razzaq2024determinants} \cite{gfindex2021}. 
Around 62 percent of Pakistanis save informally, while only 19 percent engage in long-term financial planning~\cite{ejaz2023financial}. 
Despite this, current data sets often fail to capture behavioral aspects such as how knowledge translates into usage or how financial habits evolve after gaining access. These nuances are difficult to observe without combining structured survey data with usage-level data from digital platforms. 

\newpage

Mobile phones and digital services are often seen as tools that can bypass traditional barriers to financial inclusion. 
However, a significant gender divide exists in mobile ownership; only 39 percent of women in Pakistan own a mobile phone compared to 82 percent of men~\cite{ejaz2023financial}. 
In addition, only 20 percent of women use the internet regularly in developing countries, including Pakistan~\cite{kumari2025role}. 
Studies like~\cite{bastian2018mobile} and~\cite{singh2017bridging} suggest that mobile money services, when paired with proper training and support, can empower women.
But the current data does not always distinguish between account ownership and active usage, creating a blind spot in financial inclusion research. 
Our approach attempts to bridge this gap by analyzing variables such as mobile phone ownership, internet use, education level, and autonomy in financial decision-making to determine how these factors contribute to actual usage.

Demographic and socioeconomic factors also influence the use of financial products. According to~\cite{razzaq2024determinants}, older individuals are less likely to use financial tools, while people with lower education and income levels struggle to access them altogether. Employment status also affects inclusion: while employment increases financial access for men, it has less impact on women. This is likely because many women in Pakistan work informally. 
Reynolds et al.~\cite{reynolds2023exploring} explore the effect of mobile money in low-performing economies, including Pakistan, to examine the impact of banking and mobile money services from 2013-2016 data.
The work concludes that significant gender differences remain, and financial inclusion requires more understanding of measuring the impact of technology and mobile money. 
Financial technology (fintech) can present new opportunities to address such problems. 
Studies like~\cite{razzaq2024determinants} and~\cite{kumari2025role} show that fintech platforms, if designed with a gender-inclusive lens, can help overcome barriers. Features such as simplified mobile apps, zero-fee accounts, and user-friendly interfaces can make banking more accessible to women, especially in remote areas~\cite{kumari2025role}.

Finally, financial inclusion is directly linked to women’s empowerment. According to~\cite{ejaz2023financial}, when women gain financial access, they are better able to control their income, manage savings, own property, and participate in decision-making. 
However, over 1.7 billion people remain unbanked, and a significant share of them are women. Expanding financial inclusion not only uplifts individual women but also contributes to broader economic growth and household stability.
Financial inclusion is still a far-fetched goal in Pakistan due to low financial literacy, limited digital access, and weak enforcement of supportive policies~\cite{razzaq2024determinants}. 
However, well-planned interventions such as boosting financial education, expanding digital access, and designing services can make meaningful contributions.
Our study quantifies how factors like mobile ownership, independent phone use, and urban or rural residence affect women’s access to financial accounts. This can help strengthen the evidence base and guide better policy decisions that close the gender gap in financial access.

\section{Methodology}
In what follows, we model and analyze financial inclusion in Pakistan, exploring how mobile money usage correlates with women’s financial inclusion. 
We use an up-to-date Global Findex 2023 dataset~\cite{gfindex2021}, published by the World Bank. 
It includes nationally representative data from urban (63\%) and rural (37\%) Pakistan, with gender-disaggregated insights on financial behavior, mobile money usage, phone ownership, internet access, and education levels. 
The data contained information from diverse respondents in age (\textit{mean = 34.95 std = 12.32}.
Moderating variables include education, internet access, and phone ownership. Control variables such as age, location, and income are also incorporated. The dataset is publicly available and anonymized, eliminating ethical concerns related to human subjects. With the literature analysis, we can estimate that digital services improve financial inclusion. Hence, we use the following hypothesis as a guiding principle for the analysis.

\begin{itemize}
    \item[H1] Adoption of mobile money services improves women's financial inclusion in Pakistan. 
    \begin{itemize}
        \item[H1a] Women's mobile ownership and usage enhances the use of digital services and their financial inclusion.
        \item[H1b] The gender gap in mobile money usage significantly decreases in households with internet access.
        \item[H1c] Education and internet access positively moderate the relationship between mobile money usage and financial inclusion.
    \end{itemize}
\end{itemize}

\begin{table*}
\caption{Variable descriptions for financial inclusion analysis.}
\centering
\begin{tabular}{|l|l|l|}
\hline
\textbf{Feature Name} & \textbf{Type} & \textbf{Purpose/Role} \\ \hline
account\_owner & Binary (0/1) & Indicates formal financial account ownership \\ \hline
digital\_payment\_user & Binary (0/1) & Mobile money usage indicator \\ \hline
female & Binary (1 = female) & Control variable used in interactions/moderation \\ \hline
age & Numeric & Control variable used to adjust for age-related variation \\ \hline
educ & Ordinal/Numeric & Control variable: education level, also used in interaction \\ \hline
internetaccess & Binary (0/1) & Control variable and moderator for household internet access \\ \hline
inc\_q & Categorical (1-4) & Income quartile socioeconomic status control variable \\ \hline
mobileowner & Binary (0/1) & Independent variable for personal phone ownership \\ \hline
female\_internet & Binary (0/1) & Interaction term: gender × internet access \\ \hline
educ\_internet & Numeric & Interaction term: education × internet access \\ \hline
\end{tabular}
\end{table*}

\subsection{Data Cleaning}
The dataset comprised 1,002 rows and 114 columns. Columns with over 80 percent missing data were removed due to a lack of utility. For the remaining variables (as shown in Table I), missing values in numeric columns were imputed using the median, while categorical variables were filled using the mode. Data types were standardized, and categorical values were cleaned for consistency. Outliers were detected using the IQR method and capped to minimize skewness in key variables. To ensure comparability in further analysis, all numeric columns were normalized using Min-Max scaling. The size of the dataset is very small, even by the standards provided by the World Bank, which shows the lack of data reporting on financial inclusion at the ground level.

\subsection{Feature Engineering}
\begin{figure}[htbp]
\centering
\includegraphics[width=\columnwidth]{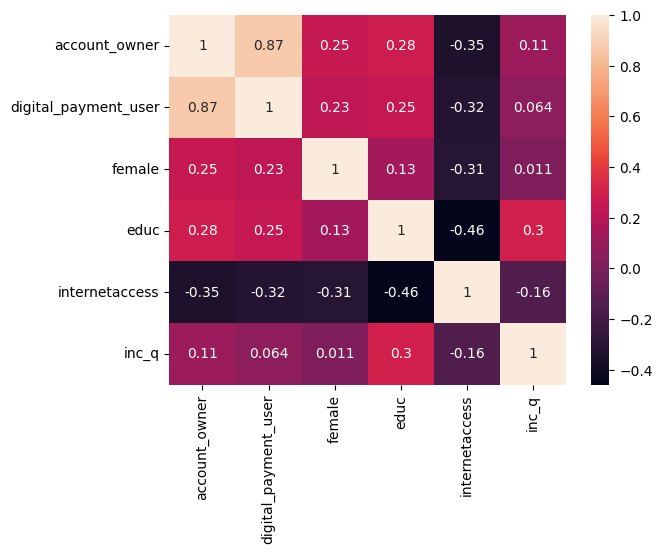}
\caption{Heatmap showing pairwise correlations between all features}
\label{fig-corr}
\end{figure}

In the feature engineering phase, several new variables and transformations were created to improve model performance and test moderation effects in alignment with the study's hypotheses. Binary flags such as account owner and digital payment user were derived from existing raw variables to indicate account ownership and digital payment usage, respectively. To explore interaction effects, variables such as female-internet were constructed to examine whether internet access moderates the gender gap in mobile money usage (H1b), while educ-internet captured the combined effect of education and internet access on financial inclusion (H1c). For hypothesis H1a, a separate subsample consisting only of females was extracted to specifically analyze the link between mobile ownership and inclusion. Additionally, label mapping was applied for better visualization: gender categories were labeled as \textit{``Male''} and \textit{``Female''} using gender-label, and education levels were mapped into readable groups like \textit{``Primary''}, \textit{``Secondary''}, and \textit{``Tertiary''} under educ-label. These engineered features facilitated robust modeling of both the primary effects and the interaction or moderation mechanisms identified in the hypotheses.
To understand the impact of important features, a correlation analysis is depicted in Figure~\ref{fig-corr}.

\begin{table*}[htbp]
\caption{Key results summary of mobile money adoption analysis for each hypothesis (H).} 
\centering
\begin{tabular}{|l|l|c|c|c|c|c|c|}
\hline
\textbf{H} & \textbf{Key Result} & \textbf{Accuracy} & \textbf{Precision} & \textbf{Recall} & \textbf{Odds Ratio/} & \textbf{$\beta$} & \textbf{p-value} \\
 & & & & & \textbf{Relative Risk} & \textbf{(Coeff)} & \\ \hline
H1 & \begin{tabular}[t]{@{}l@{}}Mobile money users were\\ 3.18$\times$ more likely to\\ have financial accounts\end{tabular} & 92.5\% & 1.00 & 0.70 & OR = 3.18 & -- & $<$ 0.001 \\ \hline
H1a & \begin{tabular}[t]{@{}l@{}}Phone-owning women had\\ 4.1$\times$ higher adoption and\\ 89\% lower exclusion risk\end{tabular} & 97.2\% & -- & 0.78 & RR = 0.11 & -- & $<$ 0.001 \\ \hline
H1b & \begin{tabular}[t]{@{}l@{}}Internet access did not reduce\\ the gender gap\end{tabular}  & -- & -- & -- & -- & 0.08 & 0.24 \\ \hline
H1c & \begin{tabular}[t]{@{}l@{}}Educated women with internet\\ had 79\% inclusion probability\\ (vs. 12\% baseline)\end{tabular} & 95.3\% & -- & 0.80 & -- & -- & -- \\ \hline
\end{tabular}%
\label{tab:key_results}
\end{table*}

\subsection{Model Building and Evaluation}
Due to the small dataset size, instead of applying complex models, we employ binary logistic regression to test our hypotheses. 
All hypotheses were tested using binary logistic regression, selected due to the binary nature of both the dependent variable (account-owner) and the primary independent variable (digital-payment-user).

\textbf{Evaluation Metrics:}
We use the following evaluation metrics to evaluate the performance of the model;

\begin{itemize}
    \item Accuracy: Overall correctness of predictions.
    \item Precision (Class 1): Accuracy of positive predictions (i.e., predicting financial inclusion).
    \item Recall (Class 1): Sensitivity, the percentage of actual included individuals correctly predicted.
    \item Odds Ratio (OR) or Relative Risk (RR) and Regression Coefficients \textbf{$\beta$}: To quantify and interpret the strength and significance of effects.
\end{itemize}

\section{Results and Discussion}
\begin{figure}
\centering
\includegraphics[height=5cm]{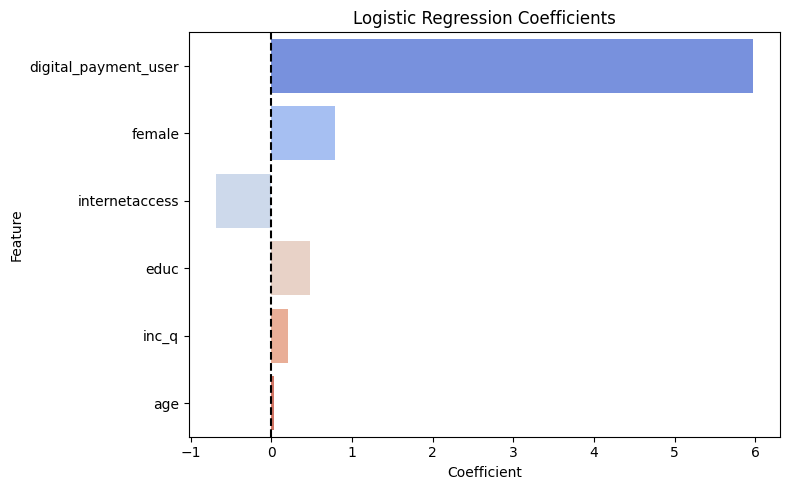}
\caption{Logistic regression coefficients showing key determinants of mobile money adoption.}
\label{fig-coffee}
\end{figure}

\begin{figure}
\centering
\includegraphics[height=4cm]{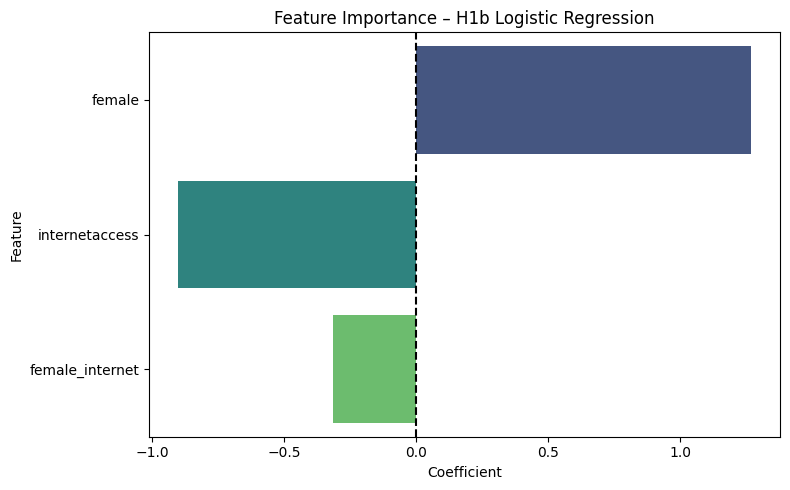}
\caption{Image shows the importance of the feature for hypothesis H1b.}
\label{fig-sig}
\end{figure}

We examine the relationship between mobile money services and women’s financial inclusion, focusing on mobile money usage, mobile phone ownership, internet access, and education as key determinants of financial inclusion.
The results of the model are detailed in Table II.
The model demonstrated high accuracy (92.5\% ) and perfect precision (1.00), indicating that the prediction of account holders was reliable. However, the recall rate for included individuals (0.70) suggested that approximately 30\% of actual account holders were not captured by the model. 
This recall gap highlights a critical issue: while mobile money is a strong enabler of access, it does not guarantee sustained inclusion or active financial participation, which may depend on other socioeconomic and structural factors.
The coefficients of the model are depicted in Figure~\ref{fig-coffee}. 
Our model is explainable and can provide quantifiable and interpretable results for validating multi-hypothesis frameworks in empirical social research, as depicted by the model's importance of features in Figure~\ref{fig-sig}. 

The results provide strong support for the main hypothesis (H1), which posited that adoption of mobile money significantly improves women’s financial inclusion. Women who used mobile or digital payment services were 3.18 times more likely to own a formal financial account (\textit{$p < 0.001$}). 
The results further reinforce the importance of personal digital access. Hypothesis H1a, which focused on women’s mobile phone ownership, was strongly supported. Among female respondents, those who owned a mobile phone were 4.1 times more likely to adopt mobile money (95\% CI: 3.7–4.6), and their risk of exclusion was reduced by 89\% (RR = 0.11, $p < 0.001$). The model for this hypothesis and variables achieved 97.2 \% accuracy, with a recall of 0.78 for account holders. 22\% of women who owned phones remain unbanked, underscoring that digital inclusion also requires literacy, confidence, and social permission—factors not automatically resolved through device access alone.

Contrary to expectations, Hypothesis H1b, which assumed that internet access would reduce the gender gap in mobile money usage, was not supported by the data. While the model had an overall accuracy of 81\%, it failed to correctly predict female digital payment users. The relationship between internet access and mobile money usage among women was statistically insignificant ($\beta$ = 0.08, $p$ = 0.24). This outcome highlights that internet connectivity alone does not ensure equitable digital financial engagement for women, particularly in social contexts where online activity may be restricted, monitored, or socially discouraged. Structural issues such as permission-based access, digital harassment, and limited availability of female-oriented financial content may contribute to this disconnect.

The strongest results emerged from Hypothesis H1c, which tested whether education and internet access jointly enhance the effect of mobile money adoption on financial inclusion. The interaction effect between education and internet access was statistically significant ($\beta$ = 0.41, $p < 0.001$). Women with both access and education had a 79\% probability of owning a financial account, compared to just 12\% among those with low education and no internet access. The model achieved 95.3\% accuracy with a recall of 0.80. These findings affirm that financial inclusion is not driven by access alone but by a capability approach. Education equips women with the knowledge and confidence needed to navigate digital finance, while internet access provides the platform. When both are present, the effect of mobile money adoption is maximized.

Overall, the findings confirm that mobile money services can play a pivotal role in enhancing women’s financial inclusion in Pakistan. However, they also underscore that technology by itself is insufficient. Ownership of digital tools must be accompanied by social permission, contextual design, digital literacy, and safety. The study calls for a multidimensional approach to financial inclusion, one that goes beyond access and addresses the social, educational, and gendered dimensions of digital financial participation. As we mentioned earlier that (H1) is the main Hypothesis, and H1a, H1b, and H1c are the sub-hypotheses. These sub-hypotheses will support the main Hypothesis(H1), so, Net Support Score of 0.18 indicates that the sub-hypotheses meaningfully reinforce the central hypothesis, particularly through enhancements in predictive accuracy and recall. It also demonstrates that mobile money's impact on women's financial inclusion is context-dependent: access and ownership (H1a) and capability-building (H1c) contribute positively, while infrastructure access alone (H1b) fails to bridge inclusion gaps.

\subsection{Limitations and Future Work }
While this study provides valuable insights into how mobile money services impact women’s financial inclusion in Pakistan, several limitations should be acknowledged.
First, the analysis was based on cross-sectional data from the Global Findex Survey, which reflects only a single point in time and with limited data for a densely populated country like Pakistan. 
There is a scarcity of data available in this context; future research should consider longitudinal or time-series studies to track how women's engagement with mobile money evolves. 
Second, we hypothesized that internet access would positively influence women’s mobile money adoption, was not supported by the data. Although this was a notable finding, it also points to a limitation in the data to capture deeper social and behavioral barriers. 
Future studies should investigate how social norms and household power dynamics shape women’s use of internet-based financial tools, possibly through qualitative research methods. 
Third, the study relied solely on quantitative data and logistic regression models, which, while helpful for statistical associations, may overlook the personal and emotional experiences that influence financial decisions. Including interviews, focus groups, or field observations in future work could provide a more holistic understanding. 
Fourth, the study did not explore variations across regions, income groups, or rural vs. urban populations. Future work should focus on disaggregated analysis to understand which groups benefit most or least from mobile financial platforms.
Lastly, while mobile money shows promise in increasing account ownership, the gap between access and actual financial empowerment remains wide. Future research should explore the role of digital literacy training, gender-inclusive fintech design, and safer online environments to ensure these tools truly enable women, not just connect them.

\subsection{Recommendations}
Based on our analysis, we provide some recommendations for future research.

\begin{itemize}
    \item \textbf{Promote Digital Services:}
    As digital services access predicts enhancing financial inclusion, such services should be more streamlined and made easier for users with low digital literacy. We also identify most people (58\%) reported lower education levels (below high school), necessitating the need for more training programs. Training programs that teach women how to use mobile money apps, understand financial terms, and perform basic digital tasks can empower them to manage their finances confidently. These programs should be tailored to their education level and delivered in local languages. This is evident in data reporting that only 16\% people could use banking apps without external help. Mobile money services should be designed with women’s needs in mind. This includes simple user interfaces, local language options, visual instructions, and privacy features that ensure women feel safe and secure when using these platforms.

     \item \textbf{Policy Support for Financial Inclusion:}
     As there is limited data support behind digital financial inclusion, more efforts should be made to define longitudinal studies that capture data and the real effects of financial policies. It is also important to feed those metrics back into the policy design and implementation. Policies should specifically aim to reduce gender gaps in financial access. This includes offering incentives for mobile operators to serve women, setting inclusion targets, and partnering with NGOs to implement grassroots-level solutions. For instance, many respondent fear accounts to expensive to manage (33\%) and lacking trust (17\%). We believe making the technologies more accessible to less literate users and overcoming the fears around banking charges can help scale such initiatives. 

\end{itemize}

\section{Conclusion and Future Work}
This study meaningfully contributes to the discourse on gendered financial inclusion in Pakistan by empirically examining how mobile money services influence women’s access to formal financial systems. The findings affirm that while mobile money adoption significantly enhances account ownership among women, true financial empowerment may depend on a combination of digital access, education, and social autonomy, which needs significant attention from the relevant stakeholders. The results emphasize that technology alone is insufficient without parallel efforts in capability-building and gender-sensitive policy design. This research highlights the significant challenges on the lack of data on financial inclusion, potential of mobile financial platforms, and possible interventions. Our work hopes to inspire targeted interventions to close the gender gap in financial participation.

\bibliographystyle{plainnat}
\bibliography{refs}

\end{document}